# Guidewire-driven deployment of high density ECoG arrays for large area brain-computer interface

Tao Zou[1,2], Na Xiao[2], Ruihong Weng[2], Yifan Guo[1,2], Danny Tat Ming Chan[3], Gilberto Ka Kit Leung[4], and Paddy Kwok Leung Chan[1,2*]

[1*]Department of Mechanical Engineering, The University of Hong Kong, Hong Kong.

[2] Advanced Biomedical Instrumentation Centre Limited, Hong Kong Science Park, Hong Kong.

[3] Neurosurgery Division, Department of Surgery, The Chinese University of Hong Kong, Hong Kong.

[4] Department of Surgery, The University of Hong Kong, Hong Kong.

*Corresponding author(s). E-mail(s): pklc@hku.hk;

**Abstract**

Electrocorticographic brain-computer interfaces (ECoG-BCIs) are powerful emergent technologies for advancing basic neuroscience research and targeted clinical interventions. However, existing devices require trade-offs between coverage area, electrode density, surgical invasiveness and complication risk – limitations that fail to meet the demands of next-generation BCI. Here, we report a guidewire-driven deployable ECoG-BCI device that can be epidurally implanted using minimally invasive procedures. Our ultra-flexible but strong thin-film electrode array, which packs 256 electrodes into 4 $cm^2$, can be folded, pulled through millimetre-sized skull holes, and unfurled seamlessly onto the brain dura mater. When deployed on the canine brain, it captures abundant high-quality auditory neural signals with distinct features of hearing that can be used to classify sound types with ≥ 80% accuracy using various standard machine learning models. Our device is biocompatible for chronic monitoring, easy and fast to deploy and importantly, resolves the key trade-offs limiting current BCI technologies.

Brain-computer interface (BCI) technologies enable the brain to communicate directly with an external device, bypassing traditional pathways like muscle and speech[1,2]. This unprecedented access to real-time brain activity has allowed us to decode how the brain functions[3-5], detect and treat neurological disorders such as epilepsy and Parkinson's disease[6-8], and even allow thoughts to control machines and restore lost functions[9-12]. Amongst the various BCI technologies, electrocorticographic BCI (ECoG-BCI), which harvests neural signals from the brain's surface[13,14], is becoming the gold standard for collecting clinical data and researching neuroscience in human subjects[15-17] because it captures better quality signals than scalp electroencephalography and is less invasive than intracortical implants that penetrate the brain tissue.

Accurately decoding brain states – whether it is processing a movement, sound, emotion or making decisions – depends on capturing fine-grained neural signals, particularly in the high-gamma (HG) frequency band (80-150 Hz). HG signals, which capture the combined electrical activity of several neurons in a small, localised area of the brain, are preferred for long-term BCI investigations because they are more stable over time than single-neuron recordings[17-20]. Several studies have shown that high density electrode arrays with high spatial resolution are needed to decode these signals effectively. During speech production, for instance, distinct HG signals ($r = 0.2$-$0.4$) in the sensorimotor cortex detected from electrodes spaced 4 mm apart revealed that the brain encodes movements related to speech in spatially discrete regions that are just millimetres wide[21]. In another study on decoding finger stimulation, smaller (0.15 $mm^2$) electrodes were found to be two times more effective at predicting how strongly a finger was stimulated than larger (15.7 $mm^2$) electrodes[22]. Micro-ECoG also consistently outperforms standard clinical recording by 36% on average in speech decoding[21]. For tasks like mapping epileptic zones[23] or studying the coupling between different cortical areas[24], electrodes that cover a large area is further needed to fully resolve the HG activity across brain networks. Emerging evidence also suggests that neural representations – the way information such as sights, sounds, thoughts and movements are stored and processed in the brain – are organized into spatially discrete millimetre scale hubs or clusters[5,25]. Targeting, sampling and decoding these information-dense neural hubs will require electrode arrays that can resolve these fine-grained functional topographies. Conventional ECoG arrays (64-128 contacts, 4-10 mm spacing), which cannot fully access high-quality HG signals, are under sampling these critical neural hubs.

Newer micro-ECoG arrays with high density and channel counts can now resolve micro-scale neural features. For example, animal studies using electrodes spaced less than 1 mm apart have uncovered fine topologies showing how sensory and auditory information are organized and processed in the brain[5,12,22,26,27]. High-resolution ECoG arrays capable of capturing and decoding detailed signals from the brain motor cortex at millimetre-scale resolution have enabled clinicians to accurately pinpoint the source of seizures in epileptic patients[5,28] and enhance the precision and responsiveness of brain-controlled prostheses[5,29]. The problem is implanting these large, high-density arrays typically require a sizeable portion of the skull to be removed[5,30]. Such craniotomies, which expose a large area of the brain tissue to the external environment, raise the risk of brain swelling[31], inflammation[32], cortical damage[33] and infection[34]. While various minimally invasive surgical techniques such as pressure-driven actuation[35,36], syringe injection[37] and shape memory actuators[38-40] have been used epidurally and sub-durally to mitigate these risks, devices deployed using these methods so far have low channel count (< 128 channels) and electrode density (< 30 electrodes $cm^{-2}$). Critically, rigid and thick (hundreds of microns) actuators induce substantial compression and shear forces on the brain, fundamentally precluding them from conforming to delicate brain tissues. For wider adoption of BCI technologies[41,42], a minimally invasive approach for deploying high-density arrays (> 60 electrodes $cm^{-2}$) that can seamlessly cover large areas of the brain is needed.

Here, we report a guidewire-driven BCI (GD-BCI) device that can be implanted epidurally over a large area of the brain via millimetre-scale craniotomies. The device, which is an ultra-flexible, high density (64 electrode $cm^{-2}$) thin-film electrode array, can be folded, pulled through the skull holes using hypercompliant guidewires attached to the array, and fully unfurled over 4 $cm^2$ of the brain dura mater, all in just 2 hours using standard neurosurgical tools (**Fig. 1a-f** and **Supplementary Video 1**). This design eliminates complex actuators, enables intraoperative repositioning and leverages the dura mater's protective barrier to minimise brain exposure. We deployed the device over the auditory cortex of a canine model and obtained high-resolution neural signal recordings that had sufficiently distinct features for highly accurate hearing decoding using standard machine learning models. Such a large coverage of high-density electrode arrays deployed on the brain surface using a minimally invasive technique is expected to significantly improve the performance and safety of BCI technologies and enable new forms of therapies that were previously inaccessible.

## GD-BCI design, fabrication, and characterization

We used photolithography techniques to create our thin-film electrode array, which consists of polyimide insulators, a gold and chromium electrode layer, and a nickel magnetic resonance imaging (MRI) marker layer (**Fig. 1e** and **Supplementary Fig. 1, 2**). This method produces a 2 x 2 $cm^2$ array consisting of 256 electrode contacts or channels. Up to 97.7% of the electrodes had impedance values ≤ 100 kΩ at 1 kHz, indicating reliable performance and high signal quality across the array (**Fig. 1g**). The spatial resolution of these electrodes ensures high ECoG signal sampling[43] for effective speech decoding[21]. To improve conformability of the array[44], we introduced perforations into the polyimide substrate to direct cerebrospinal fluid away from the electrode sites and promote a tight and stable contact with the brain surface (**Fig. 1e**). Our electrode array is 21 μm thick and has sufficient tensile strength to overcome the various stresses during implantation.

To reliably connect our high-channel-count electrode array with the signal acquisition hardware without increasing the device footprint, we used an off-the-shelf land grid array (LGA) and the CerePort circular printed circuit board (PCB) originally designed for the Utah Array. The electrical connection part of our array is manufactured in an LGA layout that bonds precisely to the connection pads of the round CerePort PCB. To ensure there is enough manoeuvre room during surgery, we added an interconnect extension to increase the separation between the electrode array and the pedestal (**Fig. 1d** and **Supplementary Fig. 3**). Because commercial skull-mounted pedestals typically have a flat base that does not conform to the skull's curvature and are constructed from heavy titanium, we designed and 3D-printed a lightweight, medical grade resin pedestal tailored to the patient's skull anatomy using computerized tomography (CT) scans (**Supplementary Fig. 4**). Our pedestal prevents detachment, does not generate artifacts, and is fully compatible with MRI.

Because the epidural space is very narrow and curved, we assessed the structural resilience of GD-BCI by applying a force on its edge to induce a shape change. The device withstood a maximum mechanical force between 1.70 and 1.90 N, which falls within the tactile precision range of neurosurgeons (**Supplementary Fig. 8**). Routine manipulation of delicate neural tissues like dissection and coagulation uses mean peak forces between 0.10 to 1.35 N[45]. As predicted by mechanical simulations, the failure initiation site with the highest stress concentration is at the junction between the guidewires and the edge of the thin film electrode array (**Supplementary Fig. 8c** and **Supplementary Fig. 9**). This design ensures that the central region, where the electrode pads are located, remains intact and stable during implantation.

Deploying the GD-BCI device on an artificial brain model using the minimally invasive procedure did not significantly affect the electrodes (**Fig. 1f** and **Supplementary Fig. 6**). The strain and deformation stress from folding and unfurling the electrode raised the impedance to > 1 MΩ for just 1.95% of the channels (initial impedance is < 1 MΩ at 1 kHz) (**Fig. 1g**). Around 93.4% of the electrodes maintained a decent impedance of < 100 kΩ and less than 50% changed their impedance after aging in phosphate buffered saline (PBS) for 7 days at 60 °C (**Fig. 1h**). Such good preservation of function is critical for chronic and large-area ECoG mapping.

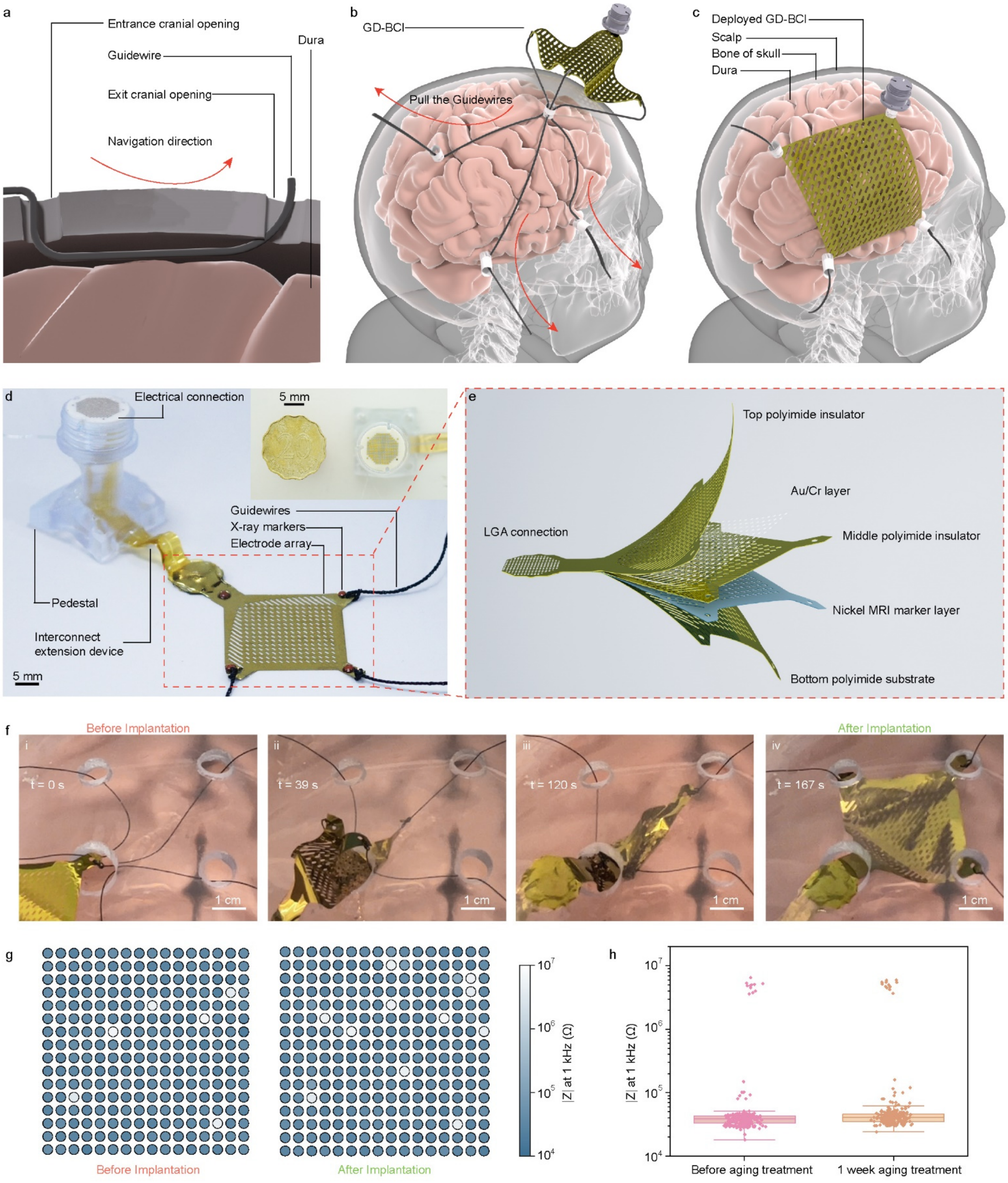

**Fig. 1: Design and characterization of GD-BCI.** **a**, Schematic of a guidewire being steered into the epidural space through two intracranial openings. **b**, **c**, Schematics showing how a folded GD-BCI is pulled (red arrows in **b**) through one of four millimetre-sized skull openings (**b**) and unfurled onto the cortex surface by mechanically retracting three guidewires (**c**). **d**, Photo of an assembled GD-BCI complete with a customized cranial pedestal that is compatible with the Blackrock Cereport headstage. Three guidewires on the edge help steer the device into the space between the skull and dura mater. Inset compares the electrical connection part of our array (right) to a 2 HKD coin (left). **e**, An exploded diagram of the thin film electrode array architecture. Total sensing coverage area: 2 cm x 2 cm, 256 channels. **f**, Photographs showing a GD-BCI being deployed onto a phantom brain model. **g**, *In vitro* impedance map of a GD-BCI with 500 μm diameter active sites before and after one implantation cycle involving a compression and expansion, measured at 1 kHz. **h**, Boxplots (n = 256 from 1 device) showing the distribution of the 1 kHz-impedance values measured before and after aging the device in PBS for 7 days at 60 °C. Median and quartile range are shown. Whiskers denote 1.5× the interquartile range. Individual data points are overlaid on the box plots.

## Virtual planning of GD-BCI localization

Precise placement of GD-BCI along the brain's intricate landscape of folded grooves (sulci) and ridges (gyri) is essential for accurately interpreting the captured neural signals. With traditional ECoG arrays that are implanted via craniotomy or durotomy, intraoperative stereophotogrammetry – an imaging technique that captures precise 3D spatial information of anatomical structures – is typically used to visually confirm the electrode placement relative to the cortical vasculature patterns with < 2 mm surface registration error using visible fiducial markers[46]. Because GD-BCI is designed for implantation without craniotomy, we used a multi-stage validation protocol to precisely position the device on the brain of our beagle dog test subjects.

For the preoperative virtual planning stage, we created a custom drilling mould to precisely locate the positions of the four skull openings. Adapting the stereoelectroencephalography (sEEG) trajectory planning[47] technique used to guide the placement of electrodes in the brain, we build 3D model of the skull and brain by co-registering CT images of the bony anatomy with MRI images of the soft tissue structures and pathology. This co-registration is critical for aligning brain regions with skull landmarks **Fig. 2a)**. To identify the cortex region where the GD-BCI should be placed, we aligned the subject's MRI T1 images with a beagle brain atlas[48] (**Fig. 2b**). Once the GD-BCI is virtually positioned on the cortex region of interest, whether it is the part that controls auditory, motor or somatosensory functions, the four skull openings are then mapped to expose the guidewire connection points (**Fig. 2c**). To accurately drill the openings, we built a custom mould with the same four holes that match the skull curvature. Manually pressing the mould onto the skull produces holes quickly at the precise location without needing expensive neuronavigational robot or intraoperative adjustments (**Fig. 2d** and **Supplementary Fig. 10a**). X-ray fluoroscopy is used throughout the procedure to ensure the device is implanted at the correct site and deployed completely (**Supplementary Fig. S11a, b**).

After surgery, we verified the implantation with MRI. The 100 nm thick ferromagnetic nickel MRI marker layer embedded within the electrode array[40] makes the GD-BCI visible by creating susceptibility artifacts or distortions in the MRI T1-image (**Supplementary Fig. 12**). By taking sequential T1-weighted MRI scans, a 3D model of the brain's anatomy along with the implanted electrode array is obtained. Co-registering the individual scans with a beagle brain population template – built from MRI scans of multiple beagle dogs – allows us to pinpoint the position of the GD-BCI device relative to standardized brain regions provided by the beagle cortical reference atlas (**Fig. 2f**).

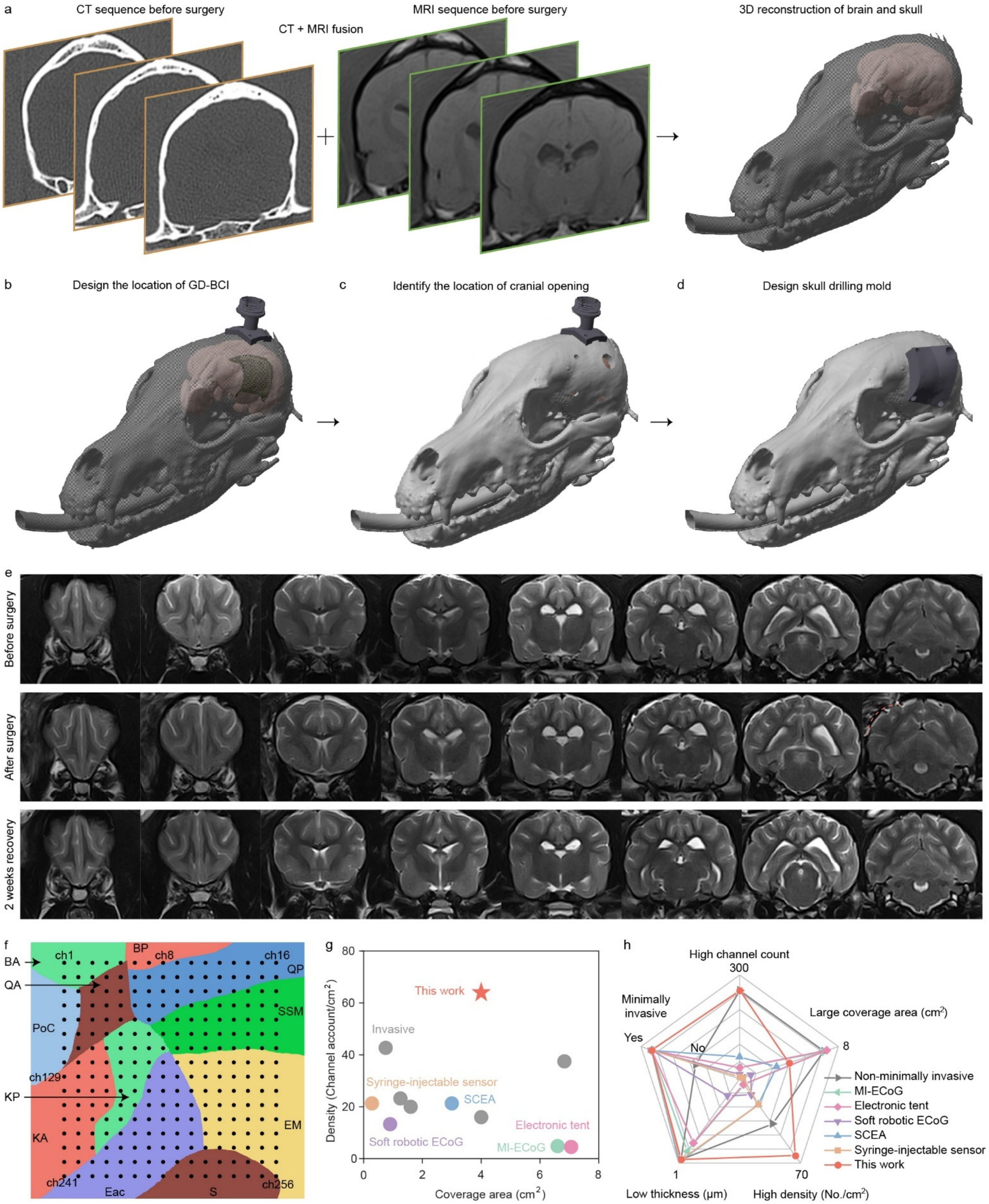

**Fig. 2: Preoperative virtual planning of GD-BCI on a beagle brain.** **a**, Overlaying preoperative CT images onto T1-weighted MRI images creates a 3D image of the skull and brain of the beagle. **b**, Aligning the subject's MRI T1 images with the beagle brain atlas identifies the cortex region where the GD-BCI and custom pedestal should be placed. **c**, Millimetre-size skull openings are determined based on the location of GD-BCI on the brain. **d**, Skull drilling mould is custom designed according to the animal's skull curvature and location of the openings for a perfect fit. **e**, T2-weighted MRI images of the beagle brain at different data points after minimally invasive implantation of GD-BCI. Dotted red line indicates position of the device. **f**, Schematic of a GD-BCI array (black circles) overlaid on an adapted beagle population template[48] outlining some brain regions. BA, entolateralis anterior lateralis; BP, entolateralis posterior; PoC, postcentralis I; QA, ectolateralis anterior; QP, ectolateralis posterior; KA, coronalis anterior; KP, coronalis posterior medialis; SSM, suprasylvian medialis; Eac, ectosylvia accessoria; S, sylvia; EM, ectosylvia medialis. **g**, **h**, Comparison of our minimally invasive GD-BCI with state-of-the-art invasive[12,21,27,49,50] and minimally invasive implants (including syringe-injectable sensor[37], soft robotic ECoG[35], SCEA[40], electronic tent[39], and MI-ECoG[36]). See Supplementary Table 1 for parameters.

## Implantation of GD-BCI on beagle dogs

After successfully deploying GD-BCI on a phantom human brain and planning the implantation site virtually, we used the minimally invasive procedure to implant the device in beagle dogs. An incision is first made on the scalp to prevent brain swelling[51], and four holes are drilled on the skull to expose the dura mater. The guidewires are then introduced into the epidural space through the skull holes using a common dura dissector (**Supplementary Fig. 10b**). Pulling the guidewires deploys the GD-BCI over the cortex (**Supplementary Fig. 10c** and **Supplementary Video 1**). Because our method does not require craniotomy, the dogs recovered very quickly and BCI experiments could be conducted on fully conscious subjects after 1 day (**Supplementary Fig. 18** and **Supplementary Video 2**). The procedure, from dural incision to layered closure, takes only 2 hours. Representative anatomical T2-weighted MRI images of the beagle brain with a GD-BCI implanted in the epidural space of the left hemisphere showed negligible changes in the brain structure immediately after and 2 weeks after surgery (**Fig. 2e**). This confirms that our method is truly minimally invasive and that the device is biocompatible and suited for long-term use.

Compared to state-of-the-art invasive and minimally-invasive implants (including syringe-injectable sensor[37], soft robotic ECoG[35], shape-changing electrode array (SCEA)[40], electronic tent[39], and minimally

invasive-ECoG[36]), our GD-BCI is superior in terms of coverage area, electrode density, thickness, and reduced tissue displacement (**Fig. 2g, h** and **Supplementary Table 1**). By replacing complex actuators such as hydraulic pumps and thermal elements with standard neurological tools and sutures, GD-BCI eliminates failure-prone components while enabling intraoperative repositioning. Doing away with rigid shape-memory scaffolds and fluidic channels has also enabled GD-BCI to achieve an ultra-flexible 21-µm-thin film configuration capable of withstanding implantation-induced traction forces and conforming seamlessly to the gyral-sulcal topology of the brain without Euler buckling[44]. Such a thin film minimizes compressive forces and interfacial shear stress on the brain, enabling GD-BCI to unfurl smoothly over a large area. Thicker devices are prone to sticking, folding or damage due to friction and tend to spread out less smoothly. Such a seamless contouring of the complex brain architecture, which is key for acquiring high-quality signals, is only possible when thin-film electrode arrays are deployed using hypercompliant guidewires. Clinically, this new approach delivers high-channel-count (256) electrodes over a large area (4 $cm^2$) while miniaturizing the fixed pedestal.

**Recording spatiotemporal dynamics of hearing from dog brain**

Once implanted, we used GD-BCI to record neural activities related to hearing across large regions of the cortex with high spatiotemporal resolution and bandwidth. Unlike conventional ECoG arrays with widely spaced electrodes, our high-density, high-throughput ECoG electrode array maps activity from groups of neurons across a wide area of the brain with millimetre-level precision. Such a comprehensive mapping capability is crucial for understanding complex and specialised responses like song selectivity[52]. In our experiments, we exposed the canines to three different frequencies (100 Hz, 1000 Hz, 10000 Hz) of pure tone sound stimulus with a sinusoidal waveform. For each frequency, the stimulus lasted 500 ms each time and was delivered 100 times at 2.5 s intervals. All neural signals recorded from the 100 trials of sound exposure were segmented to 250 ms durations, time-locked and averaged to reduce noise and yield a meaningful auditory evoked potential (AEP) signal for analysis (**Fig. 3a**).

Within 200 ms of exposure to the 100 Hz sound stimulus, the averaged response (N = 100) from a representative electrode channel displayed a distinct AEP waveform from the baseline recording before exposure (**Fig. 3a**). Examining the signals from all channels in the entire array, we observe a clear spatial pattern of AEP response (**Fig. 3b**). Larger amplitudes were recorded from channels positioned within the auditory cortex region (black square in Fig. 3b). The full unsegmented waveforms show localized responses with amplitudes reaching up to 100 μV at around 50 ms after the onset of the stimulus (**Fig. 3c**). These results suggest that the brain responded very rapidly to the sound stimulus and our electrode captured this early-stage auditory signal with high sensitivity and precision.

To better understand the spatiotemporal pattern of the neural signals after stimulation, we projected the AEP voltages from all 256 channels onto the beagle brain surface (**Fig. 3d** and **Supplementary Video 3**) and obtained a series of snapshots at 12.5 ms intervals (**Fig. 3e**). The sequential frames show that voltages over the auditory cortex increased after stimulation and decreased 25 ms later. Filtering the signals from each channel into distinct frequency bands (δ, θ, α, β, γ1, γ2) using a bandpass filter and calculating their root mean square (RMS) power, we can track a wide bandwidth of neural activity, from slow waves to fast oscillations (**Fig. 3f**). As expected, the γ2 band (80-150 Hz) map – known to be highly correlated with the location where auditory cortical activation occurs[53,54] – showed the most localized response, spanning only 3 to 4 electrodes (~ 4 mm coverage). Collectively, these results demonstrate that GD-BCI, with its ability to track neural dynamics over a large area with high spatiotemporal resolution and bandwidth, is suited for investigating the cortical activation process of hearing.

Mapping the brain's responses to the other two pure tone frequencies (1000 Hz and 10000 Hz) showed a clear frequency-dependent neural activation pattern (**Supplementary Fig. 13**). The spike duration and peak-to-peak amplitude of the AEP waveform, and the location where the strongest neural activity occurred in the brain differed for each sound frequency. These distinct features can be used to classify the response to different acoustic stimulation and decode the neural activities of hearing with high spatiotemporal resolution.

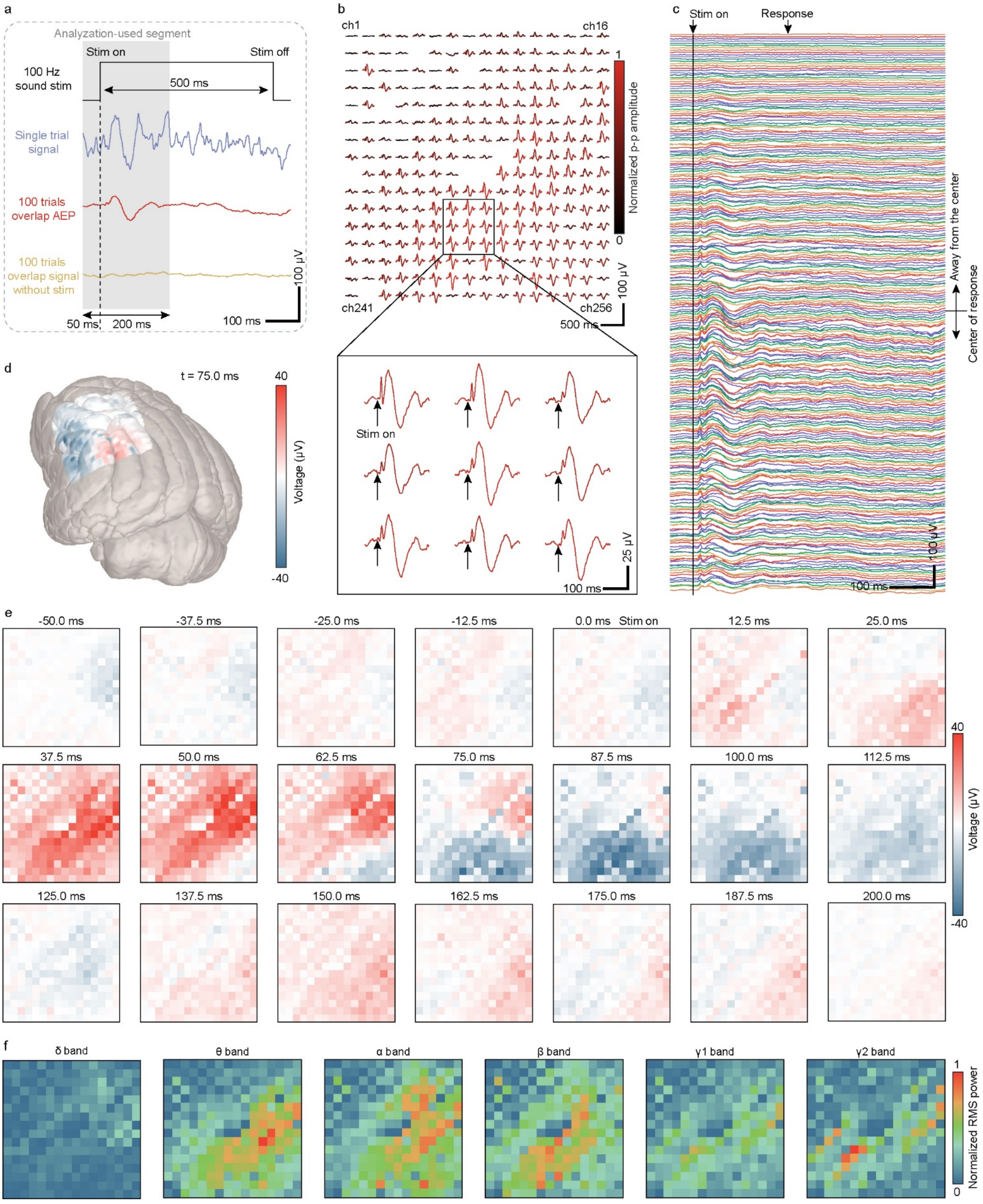

**Fig. 3: Mapping the spatiotemporal dynamics of auditory neural activity in beagles. a**, Neural signals recorded by GD-BCI. Animals were exposed to a 500 ms pure tone 100 Hz acoustic stimulus bearing a sinusoidal waveform for 100 repetitions at 2.5 s intervals. Individual recordings (blue) from the 100 repetitions are time-locked and averaged to obtain the auditory evoked potential (AEP) signal (red). For analysis, AEP signal is segmented to start at 50 ms before stimulation and end at 200 ms after stimulation (grey shaded area), giving a total duration of 250 ms. Baseline signal without stimulus (yellow) is shown for comparison. **b**, Mapped AEP response shows a clear spatial pattern. Each 250 ms segmented AEP trace corresponds to one electrode channel on the GD-BCI array. The onset of acoustic stimulation (Stim on) is marked with black arrows on signals with large amplitudes from selected channels (black square). **c**, Unsegmented AEP waveforms from all 256 electrode channels aligned to the start of the stimulus (Stim on). **d**, AEP voltage map at 75 ms after stimulation overlaid on a beagle brain surface. **e**, Movie frames at different time points taken at 12.5 ms intervals show spatiotemporal changes in the AEP voltage map. Colour scale is saturated at −40 µV and 40 µV for visual clarity. Channel 1 is at the top left corner and channel 256 at bottom right. **f**, Spatial RMS power map of the AEP filtered according to different frequency windows (δ: 1-4 Hz; θ: 4-8 Hz; α: 8-12 Hz; β: 12-30 Hz; γ1: 30-80 Hz, and γ2: 80-150 Hz).

### Decoding beagle dog hearing with machine learning

Decoding the read-outs from the auditory cortex could improve the prostheses used to stimulate parts of the central auditory pathway such as the auditory brainstem or midbrain. These types of prostheses are particularly useful for patients who cannot use cochlear implants due to damage or absence of the cochlea or cochlea nerve[12]. When these patients are asked to imagine hearing a word, the spatiotemporal dynamics from the auditory cortex can be used to decode the word and restore speech[55,56]. To decode the dog's neural responses to the different sound frequencies in our experiments, we developed a machine learning (ML) pipeline to extract features from the neural signals and deconvolute their connections with the different sound frequencies (**Fig. 4a**). Signals from individual trials across all 256 electrodes channels were segmented into 250 ms durations (Fig. 3a and Fig. 4a) and bandpass filtered between 0.5 Hz and 150 Hz. Common Spatial Pattern (CSP) analysis was applied to the filtered signals and 60 discriminative features were extracted. Each feature set was labelled according to the corresponding sound stimulus (100 Hz, 1000 Hz or 10000 Hz) and used to train a supervised Support Vector Machine (SVM) classifier. The dataset, comprising 256 electrode channels, 3 sound types, 100 trials per sound type, and 500 data points per electrode (250 ms segmented waveform x 2 kHz sampling rate), is divided using stratified random sampling to ensure each sound type is equally represented in the training and testing sets. Out of the 100 trials for each sound type, 80 trials (totalling 240 trials for all three sound types or 80% of the dataset) is allocated for classifier training. The remaining 20 trials (totalling 60 trials or 20% of the dataset) is allocated for testing.

We validated the extracted feature set by projecting the 60-dimensional space into a 3D space using principal component analysis (PCA) embedding. The distinct clusters seen in the PCA plot indicate the GD-BCI electrode captured diverse and discriminative neural signals (**Fig. 4b**). Using the noisy waveform from one trial, our classification model correctly predicted the sound type with 80.3% accuracy (**Fig. 4c**). We further tested and found that the extracted features are useful across several standard ML models, including ridge regression, SVM with linear or radial basis function kernels, and random forest classifiers. The micro-average receiver operating characteristic (ROC) curves with a 0.5 decision threshold showed that all models consistently classified the sound stimuli more effectively than random guessing (**Fig. 4d**), with median accuracy around 80% (one-way ANOVA, $p = 0.83688$, **Fig. 4e**).

Choosing the right length of the neural signal (or time window) for analysis is critical for rapid response in real-time decoding[17]. A time window that is too short might miss important neural patterns while too long adds unnecessary data and slows down the process. To find the optimal window that balances speed and accuracy, we calculated the decoding accuracy across varying time windows. The decoding performance improved with longer time windows before reaching a plateau at approximately 150 ms (**Fig. 4f**). The plateau effect was statistically confirmed by one-way ANOVA, which revealed a significant effect for time-windows between 25 ms and 50 ms ($p = 0.00101$) but not for those between 150 ms and 250 ms ($p = 0.86288$). Post hoc t-tests with Bonferroni correction indicated significant differences ($p < 0.05$) in mean accuracy across the 25-150 ms range. These results suggest that the optimal time window is around 150 ms.

To further investigate which components of the brain's frequency spectrum contributed most significantly to decoding performance, we calculated the classification accuracy using neural signals that were bandpass-filtered into distinct frequency bands (**Fig. 4g**). When using only $\gamma 2$ band signals (79.7%), decoding accuracy was not statistically different from that achieved with full-band signals (80.3%), as confirmed by paired samples t-test ($p = 0.84028$). This indicates that HG oscillations alone are sufficient for accurate classification of sound. On the contrary, the mean decoding accuracy using low-frequency bands ($\delta$ to $\beta$) showed significant differences from full-band signals (paired samples t-test). This implies that while lower frequency signals may contribute to some aspects of auditory processing such as envelope tracking or rhythmic encoding, they lack the information density or specificity present in the $\gamma$ bands that are required for fine-grained discrimination of sound type. Collectively, these results highlight the distinct roles that different brainwave frequencies play in auditory processing. Gamma oscillations, particularly within the $\gamma 2$ range, are essential for high-fidelity sound identity decoding. These insights suggest that future auditory prostheses could prioritize $\gamma 2$-band acquisition to reduce computational load, power consumption and signal processing complexity without compromising decoding accuracy.

We further used GD-BCI's high-density recording map to identify the electrodes that yielded the highest decoding performance. These spatial maps of high-performing electrodes reveal in detail the brain regions that are responding to sound stimuli – an aspect that has remained poorly characterized up until now due to the sparse coverage of standard electrode arrays. We assessed decoding performance at the electrode level using the ROC area under the curve (ROC-AUC) – a metric that captures a classifier's ability to distinguish between categories across all possible decision thresholds. The AUC value ranges from 0.5 (chance-level performance) to 1.0 (perfect discrimination). Electrodes with higher AUC contain more neural information for decoding sound types and are thus more discriminative. The heatmap shows distinct patterns associated with the classification of auditory stimulus (**Fig. 4h**). Univariate decoding performance – the ability to classify auditory stimulus using signals from a single electrode – varied significantly across the GD-BCI electrode array. Electrodes that were best at classifying specific auditory stimuli (top 10% AUC values, denoted by black borders) were spatially concentrated in the anatomically defined auditory cortex – the area below the grey line in Fig. 4h. The calculated weighted centroids (blue crosses in Fig. 4h), which represent the spatial average of the top-contributing electrodes weighted by their individual AUC values, confirm that these electrodes are clustered within the auditory cortex.

Our map also identified a subset of spatially clustered electrodes that performed well consistently across all three sound frequencies. The location where these universal high-performance electrodes are found likely correspond to putative neural hubs that are frequency-invariant and functionally essential for core auditory processing. In the context of auditory BCI, increasing the electrode density in these regions could help capture richer data and improve classification accuracy. Because the best performing electrodes were confined to a tight cluster of $< 3$ mm in at least one dimension (either the anterior-posterior or medial-lateral axes), future clinical systems must achieve millimetre-scale coverage to ensure these hubs are sampled. GD-BCI's ability to resolve high-resolution maps of sound-specific neural signals demonstrates that it is a powerful tool for capturing, analysing, and decoding the fine-scale functional architectures of auditory processing. The resolution achieved with GD-BCI is superior to standard sparse arrays and is essential for individualized neuroprosthetic applications that require precise neural interface targeting.

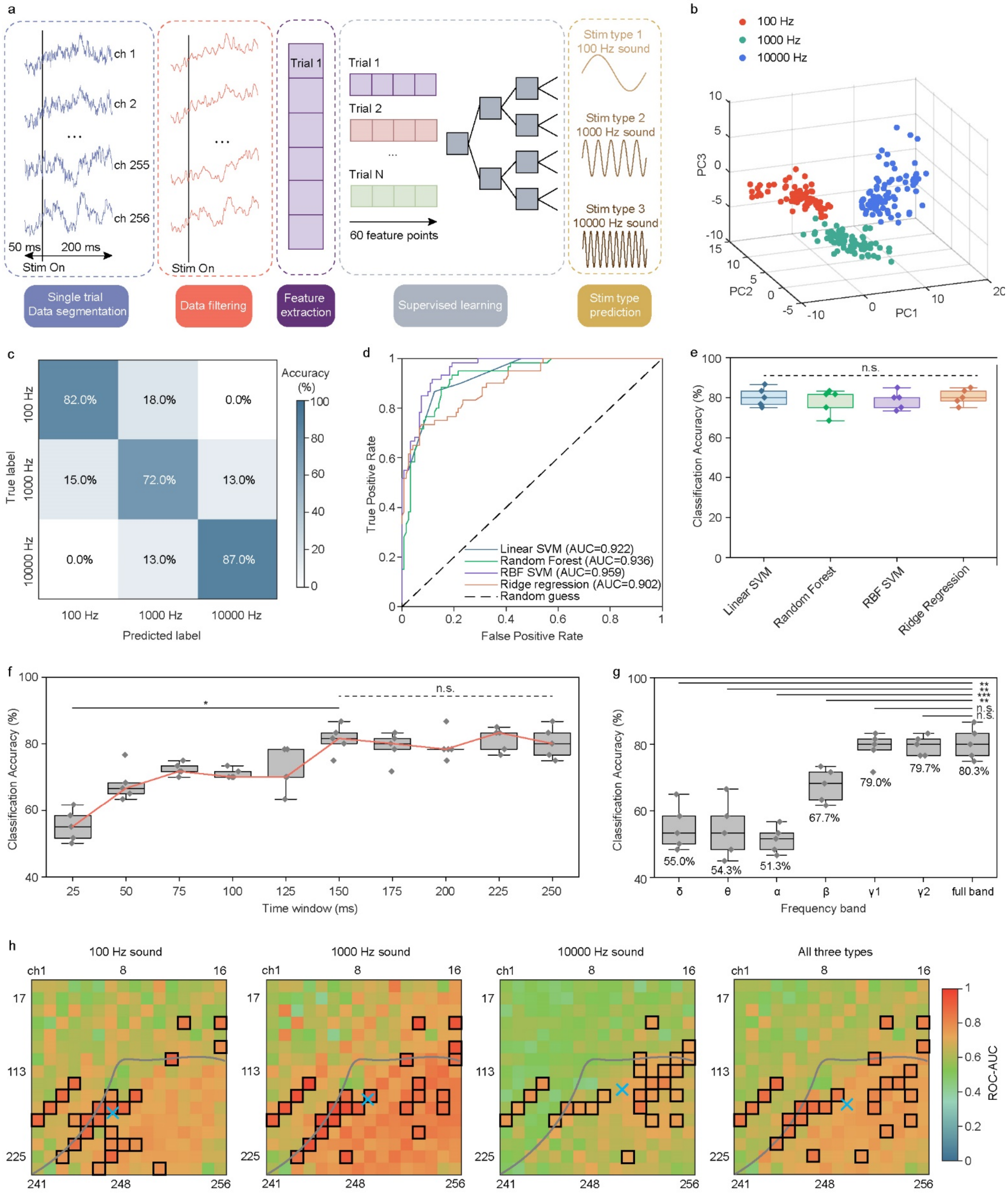

**Fig. 4: Machine learning (ML) decodes three distinct sound stimulation frequencies from neural recordings in beagle dog auditory cortex. a**, Schematic of the ML architecture for data pre-processing, feature extraction, supervised learning and prediction of acoustic stimulation type. **b**, PCA embedding plot derived from the dataset recorded by GD-BCI visually shows feature separation in 3D space. **c**, Confusion matrix displaying the classification accuracy of a linear-SVM in predicting each type of acoustic stimulation within the test set. **d**, Micro-average ROC curves of different ML models used for stimulation type classification. AUC, area under the curve. **e**, Stimulation type classification accuracy of different ML models. No significant difference is found between groups using one-way ANOVA ($p = 0.83688$). n.s., $p > 0.05$, $^{*}p < 0.05$, $^{**}p < 0.01$, $^{***}p < 0.001$, for all the subsequent hypothesis tests. **f**, Classification accuracy increases with longer time-window until saturation as validated by one-way ANOVA and post-hoc t-tests. **g**, Classification accuracy calculated using bandpass-filtered neural signals. Mean value is below each boxplot. No significant difference between $\gamma 2$ band and full band. Lower frequency bands ($\delta$ to $\beta$) were significantly less accurate than full-band signals (paired samples t-test). **h**, Spatial distribution of electrode-level decoding performance for auditory stimuli classification. Electrodes with most discriminative information for classification has the highest ROC-AUC (red). Electrodes with top 10% AUC values have black borders. Blue crosses indicate weighted centroids of high-performance electrodes (spatial average weighted by AUC values). Area below grey line is the auditory cortex. Classification accuracy in **e-g** is calculated using 5-fold cross validation ($n = 5$). Each box plot overlaid with individual data points shows the median and quartile range. Whiskers denote 1.5× the interquartile range.

### Electrode contact, density and coverage affect decoding

Given the importance of the $\gamma 2$ band activity in decoding performance, we sought to understand the spatial requirements for capturing high fidelity signals from this frequency band. Because the quality of the extracted $\gamma 2$ signals directly impacts the resolution and accuracy of neural decoding, we investigated the influence of electrode contact area (the surface area that interfaces with neural tissue) and electrode spatial density (the number of electrodes per unit area on an array) on $\gamma 2$ signal acquisition. Our results show that high spatial resolution, achieved through small electrode contact areas and high-density arrays, is essential for optimal $\gamma 2$ signal acquisition.

We assessed the impact of electrode contact size on auditory recording and decoding through an ablation study by systematically varying the design of the electrode array and measuring the impact of the changes on recording and classification accuracy. To simulate different electrode contact sizes without physically altering the electrodes, we spatially averaged raw neural recordings within virtual square subgrids ranging from 1 × 1 (effective contact size of 0.5 mm) to 8 × 8 (9.6 mm) (**Fig. 5a**). Each spatially averaged virtual electrode signal was then bandpass filtered using a $4^{th}$ order Butterworth filter (80-150 Hz) and the RMS power was computed within the 0-250 ms post stimulus window. By averaging power values across all virtual electrodes within each trial, we obtain a single γ2 band power value per trial-subgrid combination. After rescaling each γ2 band power value to a range between 0 and 1 based on the minimum and maximum power values observed across all subgrids (Min-Max normalization), we can analyse how electrode contact area affects γ2 band signal quality.

The results show that increasing electrode contact size decreases the normalised γ2 band power (Wilcoxon signed-rank test) (**Fig. 5b**). This trend, also seen in other frequency bands (**Supplementary Fig. 15**), indicates that micro-electrodes are crucial for preserving amplitude information within the γ2 band because larger contacts spatially average out and dampen the localized γ2 activity captured by smaller electrodes. Further, paired sample t-tests show that decoding accuracy decreased significantly ($p = 0.00783$) when the estimated contact size increased from 2 x 2 (effective contact size of 1.8 mm) to 4 x 4 (effective contact size of 4.5 mm) (**Fig. 5c**).

We further determined how the sound-evoked neural signals from different frequency bands are spatially distributed across the electrode array. To do this, we compared the neural signals from every electrode channel with each other for similarities by computing their correlation coefficients and displaying the values as a channel-to-channel correlation heatmap. Highly correlated electrodes indicate the neural signals have similar or overlapping activity while lowly correlated ones suggest independent or localised activity. The correlation maps show the γ2 band displayed the least correlation while the low frequency δ, θ, α, and β bands had significant inter-channel correlation (**Fig. 5d** and **Supplementary Fig. 16**). This difference arises likely because low-frequency signals tend to couple over long distances whereas high-frequency signals are more spatially constrained.

Comparing the HG envelopes – the amplitude profile of HG neural activity over time – of micro-electrode pairs using Pearson correlation further show the importance of spatial resolution and high-density electrode arrays for capturing fine-grained neural activity. As the electrode distance narrowed from 20 mm to 1.33 mm, the Pearson correlation between HG envelopes of micro-electrode pairs increased. However, at the finest electrode resolution of 1.33 mm, the signals remained largely uncorrelated ($r < 0.2$) (**Fig. 5e**) This means that as the electrodes get closer, the HG neural signals become more similar but up to a certain point. Signals from electrodes spaced 1.33 mm apart are mostly distinct. These results imply that auditory information encoded in HG neural activations is highly localised and changes over distances smaller than 2 mm. Collectively, these findings demonstrate that capturing abundant, high-quality $\gamma 2$ signals for accurate neural decoding requires electrodes with small contact size and high spatial density that can resolve fine-grained patterns of neural activity.

The impact of electrode density on decoding accuracy was investigated by down sampling the electrodes across the array to simulate lower density configurations that are 12.5%, 25%, 50% and 75% of the original full array (**Fig. 5f**). To obtain the 12.5% and 75% density configurations, we used a block-wise random down sampling method, in which the electrode array is partitioned into non-overlapping blocks and a subset of electrodes are randomly selected from each block to achieve the target density. This method ensures the down sampled electrodes are uniformly distributed across the array. For the 25% and 50% configurations, a simple checkerboard pattern is used. The results show that classification accuracy drops progressively as electrode density decreases from 100% to 75% (paired samples t-test, $p = 0.04498$) and from 50% to 12.5% (paired samples t-test, $p = 0.00614$) (**Fig. 5g**). The subsampling distribution at 25% density (equivalent to 16 electrodes $cm^{-2}$), which approximates the configuration of other minimally invasive implants (**Supplementary Table 1**), yielded significantly lower median classification accuracy (56.6%) than the full-density GD-BCI array (80.3%), as confirmed by paired samples t-tests ($p = 0.00013$). Together, these results show that to resolve $\gamma 2$ activity, next-generation auditory BCIs must adopt micro-scale electrode contacts of less than 2 mm and electrode densities that are above 60 electrodes $cm^{-2}$ like the GD-BCI.

To investigate how spatial coverage of electrodes over the auditory cortex influence decoding performance, we selected subsets of electrodes from rectangular grids spanning varying extents of the auditory cortex and examined the decoding accuracy of these different coverage areas (**Fig. 5h**). Decoding performance declined with decreasing coverage area. The median decoding accuracy showed a significant drop from 80.3% at full coverage to 54% at 25% coverage, as determined by paired samples t-test, $p = 0.0011$ (**Fig. 5i**). These findings demonstrate that sound identity is encoded across a broad region of the auditory cortex and that the broad spatial coverage provided by GD-BCI played a key role in enhancing auditory decoding under the current feature extraction and machine learning framework. Because broad spatial coverage is as important as high electrode density, arrays should prioritize large-area (> 100 $mm^2$) and high-density (> 60 electrodes $cm^{-2}$) over small, ultra-dense patches. GD-BCI, with its small electrode contact size, high spatial resolution and large spatial coverage, is well-suited for highly precise and accurate neural recording and decoding.

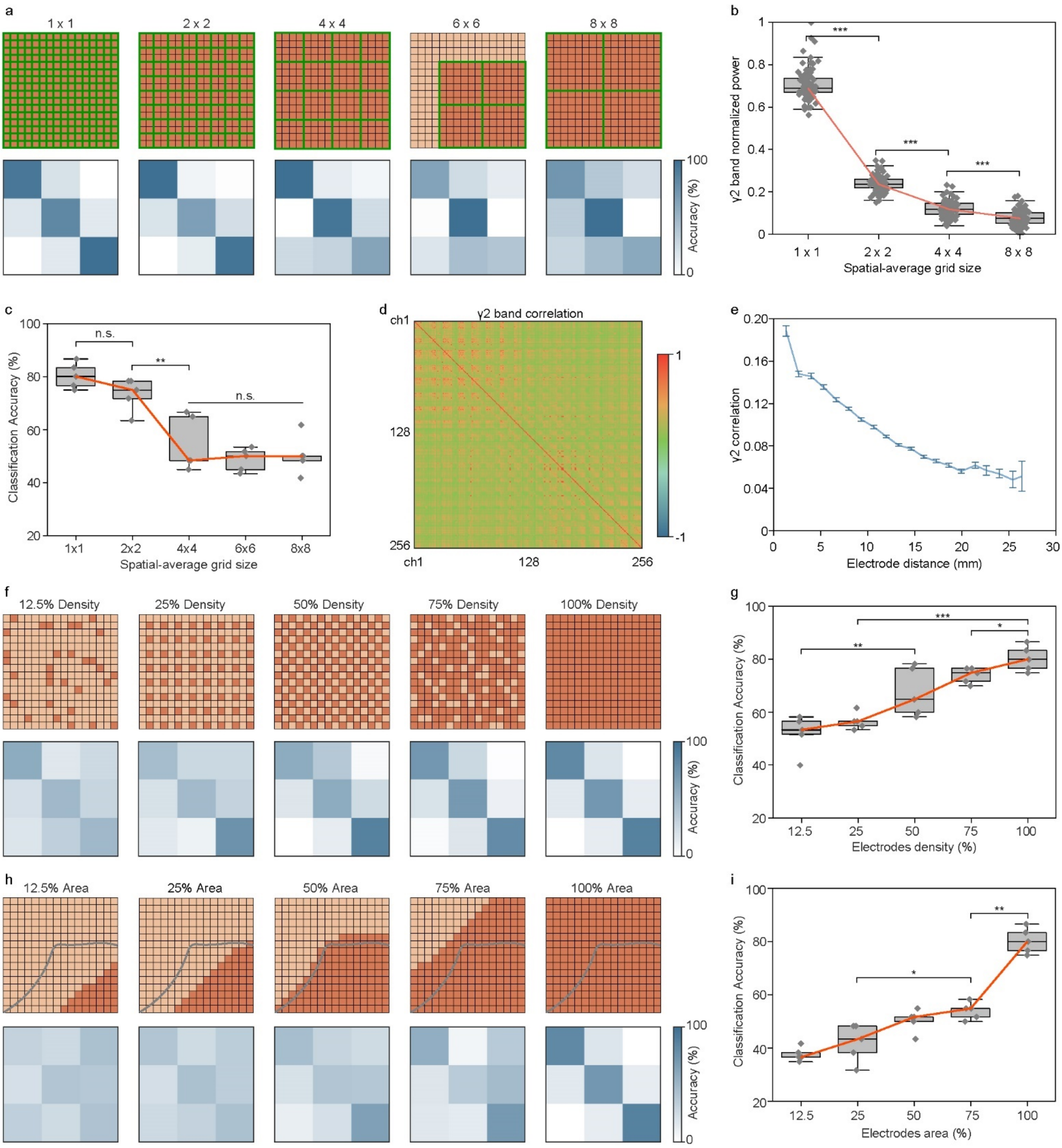

**Fig. 5: Accurate hearing decoding requires small electrode contact size, high density electrodes and large coverage areas. a**, Schematic (top row) and confusion matrices (bottom row) of different electrode contact sizes simulated by spatially averaging neural recordings within virtual square subgrids (green). **b**, Min-max normalized γ2 (80-150 Hz) band power drops significantly with electrode contact size due to

spatial averaging (Wilcoxon signed-rank test). Min-Max normalization = (Power – Min Power) / (Max Power – Min Power), where Min and Max Power are extreme band power values across all contact sizes. Boxplots show distribution across trials (n = 100). **c**, Hearing decoding accuracy decreased with increasing contact size, represented as spatially averaged electrodes as shown in **a**. **d**, Channel-to-channel correlation map of $\gamma 2$ band during sound stimulation. **e**, Inter-electrode $\gamma 2$ correlation decreased with increasing electrode distance. At spatial resolution of 1.33 mm, neural signals are distinct ($r = 0.18847$). Correlation values are mean and standard error (n = all possible electrode pairs at fixed electrode distance in mm). **f**, Schematic (top row) and confusion matrices (bottom row) of different low density electrodes simulated by spatial down sampling. **g**, Hearing decoding accuracy increased with increasing electrode density. **h**, Schematic (top row) and confusion matrices (bottom row) of different low electrode coverage area simulated by spatial cropping. Area below grey line is the auditory cortex. **i**, Hearing decoding accuracy increased with increasing electrode coverage area. Classification accuracy in **c, g, i** was calculated using 5-fold cross validation (n = 5). Each box plot is overlaid with individual data points and shows the median and quartile range. Whiskers denote 1.5× the interquartile range. Statistical comparisons are by paired samples t-test. n.s., $p > 0.05$, $^{*}p < 0.05$, $^{**}p < 0.01$, $^{***}p < 0.001$, for all the hypothesis tests.

## 3 Discussion

In this work, we have realized a minimally invasive way to implant a high-density (256 electrodes) and large-area (4 $cm^2$) ECoG-BCI device onto the brain dura mater using a guidewire-driven strategy. Our ultra-flexible but mechanically strong thin-film electrode array can be folded, pulled into narrow epidural spaces through millimetre-sized skull holes and unfurled onto the brain surface spanning 4.5 times the total skull hole areas, all in just 2 hours. We show through MRI that the brain topography and geometry of the beagle dog test subjects remained intact after implantation, allowing them to recover and move normally 1 day after surgery. Once deployed, the electrode is used to record high quality auditory neural activities, including mapping AEP signals with high spatiotemporal resolution and bandwidth. The high-density electrode configuration enabled unprecedented spatial discrimination of auditory processing in the dogs.

Our measurements revealed that the $\gamma 2$ frequency band (80-150 Hz) has highly focused cortical activation patterns, spanning just 3-4 electrodes equivalent to ~ 4 mm in size. The HG envelop correlation study further showed that neural information relevant to hearing requires sub-2 mm resolution for accurate decoding – a level of precision uniquely supported by the high spatial density of our electrode array. The inverse relationship observed between spatial resolution and signal correlation underscored the need for high-density sampling to resolve the cortical micro-architectures underlying complex auditory decoding. From the high-quality recordings obtained using our electrodes, we identified distinct electrophysiological features of hearing that could classify sound stimulation types with ≥ 80% accuracy using various standard machine learning models. Such capabilities are critical for advancing the application of neuroprostheses and deciphering the mechanism of brain networks. Up until now, similar neural signal decoding outcomes are possible only with highly invasive techniques involving large scale craniotomies[2,57,58] that are risky and difficult to heal.

Through spectral analysis, we discovered that γ2 oscillations alone are sufficient for auditory decoding, achieving similar accuracies (79.7%) to full-band signals. Our ablation studies further showed that micro-scale electrode contacts (≤ 1.8 mm), high electrode density (> 60 electrodes $cm^{-2}$) and large spatial coverage over the target auditory cortex region are all essential for capturing abundant high-quality, discriminative micro-scale neural activities that are necessary for accurate decoding. GD-BCI's decoding performance (> 80% accuracy) is attributed to its ability to resolve γ2 signals over a large area of the auditory cortex. Although the signals were segmented to 250 ms time windows in our analysis, they could, in principle, be lowered to 150 ms – the optimal time window identified from our experiments – for rapid, real-time decoding in neuroprosthetics.

By bridging minimally invasive implantation with high-fidelity neural recording and decoding, our GD-BCI technology unlocks precision neurology applications – from decoding complex brain networks to clinical treatments of diseases – without compromising surgical practicality or safety. Using surgical sutures as the guidewires and standard dural dissection procedures, our approach is fully compatible with standard neurosurgical protocols and equipment. Because deployment and extraction of the electrode array are done through the same skull holes, no additional craniotomy procedures are needed for post-monitoring. Compared to existing mass-produced implants[59], our customized 3D-printed pedestal is slim, lightweight, easy to fasten and importantly, adapts to the skull anatomy of the wearer. All these are essential features for chronic recording.  The skull drilling mould also simplifies and speeds up the localization process. Furthermore, because the deployment technique is independent of the device's size or shape, GD-BCI's geometry can be tailored for either temporary wide-area cortical mapping or long-term monitoring of specific brain regions.

In the future, our GD-BCI delivery framework could support integrated systems that combine clinically proven stimulation techniques like deep brain stimulation or cochlear implants with emerging computational hardware like AI accelerators or neuromorphic chips for decoding complex brain functions such as perception and cognition in humans. We anticipate that GD-BCI will enable new forms of bioelectronic therapies that were previously unfeasible due to limitations in surgical access.

## 4 Methods

### 4.1 Fabrication and assembly of GD-BCI

The thin film electrode array was fabricated on a 4-inch Si wafer (Thermal oxidation processing, Namkang Hi-Tech) using standards photolithography processing (**Supplementary Fig. 1**). (1) A commonly used biocompatible substrate material (polyimide, SG1020L, Shengbang polymer materials Co., Ltd.) was spin coated onto the wafer at 500/3000 rpm for 10/60 s, respectively. Subsequently, the polyimide was soft baked at 100 °C for 3 mins, followed by a curing process in a nitrogen environment at 100 °C, 150 °C, and 350 °C for 30 mins, 40 mins, and 40 mins, respectively. The polyimide thickness was 7 μm after final curing. (2) The bottom polyimide substrate was patterned by a photolithography and a reactive ion etching (RIE) process using AZ nLOF 2070 photoresist (MicroChemicals), mask aligner system (MA/BA6, Karl Suss), and plasma etcher (Oxford Plasma Pro 100 RIE). (3) The nickel MRI marker (100 nm thickness) was formed on the patterned polyimide substrate by a photolithography, metal deposition, and lift-off process using LOR 3A (MicroChem)/ S1813 (Kayaku Advanced Materials) double-layer photoresist, mask aligner system (MA/BA6, Karl Suss), sputter evaporator (Kurt J. Lesker), and lift-off remover (dimethylsulfoxide, Sigma-Aldrich). (4) The middle polyimide insulator layer was formed and patterned over the nickel layer using the same processes. (5) Metal layer consisting of Cr/Au (10 nm/100 nm) was formed by a photolithography, metal deposition, and lift-off process using LOR 3A (MicroChem)/ S1813 (Kayaku Advanced Materials) double-layer photoresist, mask aligner system (MA/BA6, Karl Suss), thermal evaporator (Kurt J. Lesker), and lift-off remover (dimethylsulfoxide, Sigma-Aldrich). This metal layer was used as recording electrodes, interconnect line, and land grid array (LGA) connection pad. (6) The top polyimide insulator layer was formed and patterned over the Cr/Au layer using the same processes. The exposed recording electrodes sites were 500 μm. (7) The fabricated thin film electrode array was detached from the Si wafer by mechanical peel off for the following assemble process. The total thickness of the thin film electrode array was 21 μm.

To minimize the overall size of overhead implants while maintaining a high channel count, the electrical connection section of the thin-film electrode array was patterned in a LGA configuration that precisely matches the connection pads of the CerePort circular PCB (Blackrock Neurotech). An interconnect extension device was fabricated as described above (without middle polyimide insulator and nickel MRI marker) to connect the thin film electrode and the CerePort round PCB (**Supplementary Fig. 3**).

To create the X-ray markers, fine copper powder (mesh 10000, Sigma-Aldrich) was homogeneously mixed into a two-part PDMS silicone (Sylgard 184, Dow). A 1:4 ratio of curing agent to silicone base was used. The powder was added at a Cu:PDMS 1:1 w/w ratio, and the mixture was combined in a planetary mixer (AR-100, Thinky) for 2 mins at 2000 rpm. Subsequently, the mixture was spin coated onto a 4-inch Si wafer at 200 rpm for 30 s and cured at 100 °C for 1 h. The thickness of the mixture film was 320 μm to be opaque enough for X-ray. The X-ray marker was cut into 1.5 mm diameter circle by a metal mould (**Fig. 1d, Supplementary Fig. 11a**).

According to the previously reported modelling procedure[60], the pedestal was custom designed to ensure a seamless fit between bone and implant, which contribute to problems with infection and implant detachment (**Fig. 1b, Fig. 2b-c, Supplementary Fig. 4**). The surface of the pedestal base was designed as same as the skull curvature of the beagle participant. The skull structure was reconstructed by CT imaging (Apsaras16, Kangda Intercontinental Medical Equipment) of the beagle participant. The pedestal utilized a "screw-on" quick interconnect to the headstage (CerePlex E256, Blackrock Neurotech). The "screw-on part" of the pedestal was designed according to the dimension of CerePort pedestal (Blackrock Neurotech). The whole pedestal was 3D-printed out of medical-grade resin (Biomed clear, Formlabs Inc.) using a desktop stereolithography printer (Form 3B+, Formlabs Inc.). To limit corrosion process and wear particle formation of the pedestal as previously suggested[61], a 5-μm-thick parylene C was deposited to the surface of the pedestal (PDS 2010, Specialty Coating Systems Inc.).

The LGA electrical connection part of the interconnect extension device was bonded on the CerePort round PCB using silver epoxy (MG Chemicals 8331S). Silver epoxy was selectively deposited on the footprints of the round PCB pad by screen printing method using a 100-μm-thick metal mask (**Supplementary Fig. 5**). The amount of silver epoxy that was required for reliable bonding was optimized by adjusting the size of the holes in the metal mask (**Supplementary Fig. 5c**). The silver epoxy before and after patterned over the PCB is shown in **Supplementary Fig. 5a-b**. A customized design micro-alignment stages with 6-axis degrees of freedom and a vacuum chuck heating table was used to precisely align the interconnect extension device and the CerePort round PCB. Once aligned and placed in contact, the interconnect extension device and the CerePort round PCB were cured on 80 °C hotplate for 2 h to fully cure the silver epoxy and ensure electrical connection across all bonding contacts. The bonding interface of the connection pads was shown in **Supplementary Fig. 5d**. The thin film electrode array and the interconnect extension device was also bonded as described above.

To deploy the GD-BCI, three 20-cm-length surgical sutures were fixed to the edges of the thin film electrodes with knots as the guidewires (Size 1, Jinhuan Medical Products Co.,Ltd.). The reference and ground wires (Silver wire with 0.008'' core, A-M Systems) were soldered over the footprints of the round PCB pad. The X-ray markers were sticked to the edge of the thin film electrode array using silicone adhesive (Kwik-sil, World Precision Instruments). The silicone adhesive was also used to insulate the bonding parts of the assembled GD-BCI. The final assembled GD-BCI was packed into a sterilized bag (Sanqiang Medical) and then sterilized using hydrogen peroxide plasma sterilization (SQ-WD-100, Sanqiang Medical) before surgery implantation.

### 4.2 Deployment demonstration in brain phantom

The brain phantom (**Fig. 1f and Supplementary Fig. 6**) was fabricated by 3D printed resin according to a open source anatomy atlas (https://www.openanatomy.org). The minimal gap between the dura mater phantom (light pink) and the head skull phantom (transparent) was 2 mm which is comparable with real human epidural space distance. To mimic the adipose tissue and connective tissue in human epidural space with Young's modulus of 1.6-5.5 KPa[62], an agarose gel was inject into the phantom gap. The agarose gel was made by mixing 0.2 wt% agarose powder (MB755-0500, Bio-Helix Co., Ltd.) in deionized water. The Young's modulus of the artificial tissue gel was measured by exerting loads on a cube gel (2 cm on each side) with a tensile-compressive tester (Zwick Roell). The load-displacement curve was translated to the stress strain curve for fitting the modulus, which is the slope of the curve (**Supplementary Fig. 7**). The skull model featured a small hole with a diameter ranging from 4 to 5 mm, achieved through drilling.

The guidewires were steered inside the phantom gap by a dura dissector. After all the guidewires were navigated to the designed position, the GD-BCI are deployed by manually pulling the guidewires.

### 4.3 In-vitro electrical performance characterization

To electrically characterise the GD-BCI, a digital headstage (CerePlex E256, Blackrock Neurotech) was connected to the device. Impedance was measured at 1 kHz, with the device soaped in phosphate-buffered saline (PBS) solution (0.01 M, Sigma-Aldrich) with a Ag/AgCl reference electrode. The two-point impedance measurements were taken before and after the implantation into the brain phantom. The aging treatment of the GD-BCI was validated by immersing the device in 60 °C of PBS for 7 days, which is equivalent to 5 weeks in 37 °C of PBS[63].

### 4.4 Localization of GD-BCI

Prior to the surgery, preoperative CT scanning and T1-weighted MRI scanning were carried out on beagle dogs. The 3D model of the skull was generated by RadiAnt DICOM Viewer. 3D reconstruction from serial MRI images was conducted after skull stripping using FSL software (FMRIB, Oxford, UK, https://fsl.fmrib.ox.ac.uk/fsl/fslwiki/FSL). The brain model and skull model was then aligned using MeshLab and 3D slicer software according to the reported process procedures[60]. To identify the regions of interest, the registration of the subject's T1 images to the atlas T1 images of beagle population template (https://ecommons.cornell.edu/handle/1813/67018)[48] was conducted with 3D slicer software. According to the atlas, the thin film electrode array was designed to cover the auditory cortex. The location of the four skull openings was then identified over the aligned model. Adapted from the guiding chamber and guiding screw used in sEEG probe alignment[47], a skull-opening drilling mould with the same four holes was developed for localization of craniotomy positions. The mould's contact surface was engineered to precisely match the skull curvature of the opening area, enabling accurate surgical positioning through manual pressure application that ensures complete mould-skull conformity. The mould was 3D printed by the same method as described in the above pedestal fabrication section.

During the surgery, the skull-opening drilling mould was placed over the exposed skull and then pushed with a certain force to achieve a perfect fit (**Supplementary Fig. 10a**). The skull holes were drilled along the axis of the four holes of the mould. The GD-BCI device was then deployed through these holes. Before finally suturing the wound, an X-ray imaging was carried out on a Numen pet DRF (Quantum Tec Medical Devices) to verify the marker position at the edge of the device (**Supplementary Fig. 11a-b**).

The T1-weighted MRI scans were done again after the GD-BCI implantation for postoperative localization and verification (**Supplementary Fig. 12**). The brain model was then reconstructed as described above. Similar to the previous report[40], device localization was enabled through MRI-visible susceptibility artifacts created by the intentionally designed ferromagnetic MRI marker, observed as consistent signal voids in T1 MRI sequence images. The serial MRI images were used to obtain 3D reconstructed model of both the brain and the thin film electrode array. After registering MRI images of population template onto the subject's brain model as described above, the exact locations of the electrode sites relative to the anatomical structures of the brain can be determined from the atlas on the template.

### 4.5 Minimally invasive implantation of GD-BCI into beagle dog and MRI studies

Four implantations of the GD-BCI were performed in beagle dogs. The canine experiments were approved by the ethics committee of Guangzhou Huateng Biomedical Technology Co., Ltd. (Guangdong, China). All the procedures were conducted in accordance with the guidelines outlined in the Association for Assessment and Accreditation of Laboratory Animal Care International (AAALAC). The devices were implanted to beagles epidurally through the minimally invasive procedure. Four adult male beagles (12-24 months, 10-15 kg) were obtained from and housed at Guangzhou Huateng Biomedical Technology Co., Ltd.. Prior to the experiments, animals were considered healthy on the basis of clinical examination. The general anaesthesia was performed by a board-certified veterinary anaesthesiologist during the whole process. Each dog was premedicated with an intravenous injection of a mixture of butorphanol (0.3 mg/kg), ceftriaxone (25 mg/kg), mannitol (2 g/kg at 20 wt%), and etamsylate (0.1 mg/kg) for deep sedation, infection prevention, intracranial pressure reduction, and intraoperative bleeding reduction. An indwelling needle was left in the cephalic vein of the left forearm for vascular access. Dogs were intubated after being induced to general anaesthesia with propofol (0.5 mg/kg). The general anaesthesia was maintained with inhalant isoflurane (1–2%) in oxygen (0.5 L/min) provided through an anaesthesia machine (WATO EX-35Vet, Mindray Animal Medical). During the surgery, constant rate infusion (CRI) of a mixture of lidocaine (0.2 mg/kg/h) and butorphanol (0.5 mg/kg/h) was used for pain management, and CRI of normal saline (2 ml/kg/h) was used to maintain fluid perfusion. Standard anaesthetic monitoring was set up by a veterinary patient monitor (BeneVision N15, Mindray Animal Medical), including heart rate, oxygen saturation, non-invasive blood pressure, respiration, rectal temperature, and capnometry ($CO_2$). The dog was maintained oxygen saturation level >92% throughout the surgery. For the entire procedure, animals were placed in sternal recumbency. Upon skull exposure, burr holes (one 8-mm-diameter hole and three 4-mm-diameter holes at four corners of a 2.5 cm by 2.5 cm square) were drilled using a pneumatic powered drill (Surgairtome Two, ConMed) equipped with a 3-mm-diameter Spherical burr. The burr holes were drilled under guidance of a custom designed drilling mould (**Fig. 2d, Supplementary Fig. 10a**). The dura matter was then detached from the skull using the dura dissector. To reduce the friction from tacky biological tissue, a lubricant solution (Sodium Hyaluronate, Ryan Biological) was injected into the epidural space. For a GD-BCI implantation, the three guidewires combined on the edge of the thin film electrode array were inserted into the largest hole. With the leading of the dura dissector, the guidewires were steered out of the other three holes one by one. After all the guidewires were navigated out of the three holes, a pulling force (no more than the maximum force, **Supplementary Fig. 8, 9**) was further applied manually on the guidewires to deploy the thin film electrode array on the epidural space. Finally, the pulling threads were cut after the probe was unfolded properly on the surface of the cortex and the pedestal was fixed over skull using titanium screw.

To evaluate the influence of the GD-BCI implantation, T2-weighted MRI studies were performed on a Superscan-1.5 (Xingaoyi Medical Equipment Co.,Ltd.) at different timepoints after the surgery. The dog was anesthetized and monitored as described above during MRI scanning. The scanning was done coronally using a standard protocol.

### 4.6 Pure tone acoustic stimulation

The pure tone acoustic stimulation paradigm used in this project followed a classical steady-state single-frequency paradigm reported previously[64] under anaesthesia status. The tone stimulation with three different frequencies (100Hz, 1000Hz, 10000Hz) was delivered by a speaker in the animal right ear under control of a microcontroller (Arduino R3). Each frequency of stimulation is consisting of 100 acoustic stimuli trials with 500 ms duration. The inter-stimulus interval was 2.5 s. Then, each trial of the recordings was aligned and averaged based on the TTL pulses time-locked to the stimulation for analysis the AEP. The AEP signals were segmented start 50 ms before stimulation and end 200 ms after stimulation for analysis.

## 4.7 Neural Recording

The beagle was medicated with a pain reliever (butorphanol, 0.3 mg/kg) before collecting the signals to ensure its comfort. The digital headstage (CerePlex E256, Blackrock Neurotech) was connected to the LGA electrical connection part positioned over the scalp for neural recording. A titanium screw near the craniotomy was used as the ground electrode, and silver wires inserted into the epidural space were used as reference electrode. All signal analysis was done using MATLAB (MathWorks). Signals were sampled at 2 kHz. Bandpass 0.5-150 Hz and 50 Hz notch filters were applied to the raw data to filter out the ECoG signal from background noise. Several channels were not shown in the AEP mapping due to the interference in some trials. The AEP activities mapping was projected over the beagle brain anatomy using Brainstorm software (https://neuroimage.usc.edu/brainstorm/)[65]. The spatiotemporal dynamics of the AEP activities were reflected in the movie frames (frame interval: 12.5 ms). The Spatial RMS power mapping filtered in different frequency bands (δ: 1-4 Hz; θ: 4-8 Hz; α: 8-12 Hz; β: 12-30 Hz; γ1: 30-80 Hz, and γ2: 80-150 Hz) was quantified within the segmented period window using four-order Butterworth bandpass filters. Spontaneous signals for 3 days after implantation were also recorded (**Supplementary Fig. 17**). Recordings were carried out on two beagle dogs with one representative example shown in this work.

## 4.8 Machine learning pipeline for hearing decoding

All data pre-processing, feature extraction, and training models building were done using MATLAB (R2021b). All single trial signals used in machine learning were also segmented start 50 ms before stimulation and end 200 ms after stimulation (in total 250 ms duration). The segmented signals were bandpass filtered from 0.5 Hz to 150 Hz. The dataset (500 data points × 256 channels × 100 trials × 3 stimulation types) was divided through stratified random splitting, allocating 240 trials (80%) for classifier training while reserving 60 trials (20%) for prediction evaluation. This partitioning strategy maintained equivalent class distributions between training and testing subsets, forming a balance dataset.

The feature extraction process was done by a pairwise common spatial pattern (CSP) approach[66,67]. For each class pair combination, spatial filtering was performed through covariance matrix optimization using training set. The trial-averaged covariance matrices from the ECoG time series were calculated by the equation (1):

$$\mathbf{\Sigma}_1 = \frac{\mathbf{X}_1\mathbf{X}_1^T}{\mathrm{tr}(\mathbf{X}_1\mathbf{X}_1^T)},\ \mathbf{\Sigma}_2 = \frac{\mathbf{X}_2\mathbf{X}_2^T}{\mathrm{tr}(\mathbf{X}_2\mathbf{X}_2^T)} \quad (1)$$

where $\mathbf{\Sigma}_1$ and $\mathbf{\Sigma}_2$ represent trial-averaged covariance matrices for classes 1 and 2 respectively, with $\mathbf{X}_1$ and $\mathbf{X}_2$ denoting the time-windowed neural signals for all trial in classes 1 and 2 respectively (500 data points × 256 channels × 80 trials in training set). The composite covariance matrix was subsequently decomposed through generalized eigenvalue analysis in equation (2):

$$\mathbf{\Sigma}_1\mathbf{W}=\lambda(\mathbf{\Sigma}_1+\mathbf{\Sigma}_2)\mathbf{W} \tag{2}$$

where the eigenvectors **W** were retained as discriminative filters using training set, corresponding to sorted eigenvalues (the largest and smallest). These spatial projections were then applied to both training set and testing set through matrix multiplication $\mathbf{W}^T\mathbf{X}_t$ (t: trial index), with logarithmic variances of the transformed signals calculated as train features and test features. The process was repeated across all three class pairs, concatenating features from each pair's extremal filters (10 largest/smallest eigenvalues per pair) to form a final 60-dimensional feature vector for each trial.

The extracted features were projected to a three-dimensional space using principal component analysis (PCA) for visualizing high-dimensional data. Several ML models were developed to classify ECoG signal features for different stimulation type, including ridge regression, SVMs with linear or radial basis function, and random forest. The ML models were evaluated according to their micro-average receiver operating characteristic curves (ROC) and classification accuracy.

The decoding strategy was used to inter the spatial representation of sound stimulation features. An SVM decoder model was trained on each electrode to decode these stimulation types. The area under the ROC Curve (ROC-AUC or AUC) was calculated for each electrode to quantify the specificity of stimulation feature encoding.

To study the effect of electrodes density and electrodes coverage area of GD-BCI on classification, the electrodes was down sampled across the array using block-wise random down sampling method. Next, to demonstrate the importance of coverage area, the electrodes were cropped from rectangular grids at fixed spatial resolutions.

The classification accuracy was verified with 5-fold cross validation (**Supplementary Fig. 14**). The data set was randomly split into 5 subsets. Prediction was performed for each subset using a decoder trained with the other 4 subsets. Accuracy was defined by (the number of correct predictions)/ (the number of all predictions).

### 4.9 Statistical analysis

Multiple statistical methods were performed to validate GD-BCI recordings for sound stimulation decoding using MATLAB (R2021b). Normality was assessed first using Kolmogorov-Smirnov tests and homogeneity of variances using Levene's test. Statistical analyses were then performed using non-parametric and parametric tests as appropriate for data distributions. To evaluate the decoding performance of different ML models, one-way ANOVA was used to estimate the significance. Time-window effects on decoding performance were evaluated with one-way ANOVA and post-hoc t-tests. Frequency-band classification accuracy was compared using paired sample t-tests between the full-band signals and the other frequency-band signals. For γ2 band power attenuation with spatial averaging, significance was assessed using the Wilcoxon signed-rank test on min-max-scaled normalized power values across trials ($n = 100$). Electrode distance effects on γ2 band correlation were quantified using Pearson correlation coefficients, with mean ± standard error reported for all possible electrode pairs at fixed distances. Decoding accuracy employed paired samples t-tests to evaluate contact size effects using spatially averaged electrodes, electrode density effects using uniformly-down sampled electrodes, and coverage area effects using spatially cropped electrodes. Significance thresholds were set at $\alpha = 0.05$ for all tests, with Bonferroni correction applied for multiple comparisons where appropriate.

## 5 Acknowledgements

This study was supported by the InnoHK initiative of the Innovation and Technology Commission of the Hong Kong Special Administrative Region Government. We thank A.L. Chun of Science Storylab for valuable discussions and for critically reading and editing the manuscript.

## 6 Author Contributions

T.Z. and P.K.L.C. conceived and designed the research. T.Z. performed the device design, fabrication, and development. T.Z. and Y.G. performed the in-vitro experiments. D.T.M.C. and G.K.K.L. provided expert clinical advice and surgical guidance regarding electrode implantation. T.Z. and N.X. carried out the in-vivo implantation experiments. T.Z., N.X., and R.W. performed the auditory stimulation experiments. T.Z. and R.W. analysed and decoded the auditory neural signals. T.Z. and P.K.L.C. wrote the manuscript with input from all authors. All authors provided input during manuscript preparation and revisions. P.K.L.C. supervised the study.

## 7 Ethics declarations

### 7.1 Competing interests

T.Z., N.X., and P.K.L.C. are inventors on a US provisional patent related to this work filed by the University of Hong Kong and Advanced Biomedical Instrumentation Centre Limited (no. 63/564,204, filed on 12 Mar 2024). T.Z. and P.K.L.C. are inventors on a US provisional patent related to this work filed by the University of Hong Kong and Advanced Biomedical Instrumentation Centre Limited (no. 63/763,609, filed on 26 Feb 2025). The remaining authors declare no conflict of interest.